# Solve the Refugee Crisis with Data

## Yunfei Liu

## Summary

In this study, we addressed the refugee crisis through two main models. For predicting the ultimate number of refugees, we first established a Logistic Regression Model, but due to the limited data points, its prediction accuracy was suboptimal. Consequently, we incorporated Gray Theory to develop the Gary Verhulst Model, which provided scientifically sound and reasonable predictions. Statistical tests comparing both models highlighted the superiority of the Gary Verhulst Model.

For formulating refugee allocation schemes, we initially used the Factor Analysis Method but found it too subjective and lacking in rigorous validation measures. We then developed a Refugee Allocation Model based on the Analytic Hierarchy Process (AHP), which absorbed the advantages of the former method. This model underwent extensive validation and passed consistency checks, resulting in an effective and scientific refugee allocation scheme. We also compared our model with the current allocation schemes and proposed improvements.

Finally, we discussed the advantages and disadvantages of our models, their applicability, and scalability. Sensitivity analysis was conducted, and directions for future model improvements were identified.

**Key words: Logistic Regression    Gary Verhulst Model    AHP    Compared**

# Catalogue

# 1. Introduction

The Syrian refugee crisis is escalating nowadays. Refugees became the most severe headache of the European Union and other Middle East countries. According to the information of the UN refugee agency (UNHCR) and the European Union statistics website, now the number of refugees around the world comes up to a total of 59.5 million, achieving the highest number since the second world war. In terms of such a huge refugee group, there is no doubt that how to settle them down is certainly a very tricky problem. According to the statistics of the international organization for migration, there are 37% of the refugees from Syria, 8% of the refugees from Eritrea, 8% from Afghanistan. Due to the geographical, economic, political and other factors, most of the refugees rushed into Europe, causing the refugee crisis in Europe.

For this matter, after many negotiations, the European foreign minister determined an acceptance scheme for the refugees from all over the world which is based on the factors such as gross domestic product (GDP), population and acceptance number of refugees. However, some European Union countries did not accept this "band-aid" solution. The opinions and solutions of every country about the refugee crisis were not exactly the same.

In this study, we aimed to address the refugee crisis in Europe by developing models to predict the overall number of refugees and to formulate an optimal refugee allocation scheme. For the first question, we established two models to better analyze the refugee conditions in Europe. Model one utilized a simple Logistic Regression Model. However, due to the limited data available (only statistical data from January to October 2015), the fitting condition was not very accurate, resulting in a larger error than anticipated. Consequently, we developed Model two based on the Gray Verhulst Model. This model provided a better prediction result through algorithm implementation. To verify the accuracy of our model, we also compared the results with the number of Iraqi refugees.

For the second question, we again developed two models to create the best refugee allocation scheme. Model one used the Factor Analysis Method, which is easy to understand but suffers from subjective parameter selection, lack of a realistic basis, and sensitivity to noise. To address these issues, we constructed an allocation model based on the Analytic Hierarchy Process (AHP). By iteratively testing values, we established an evaluation matrix based on criteria such as economic conditions and geographical factors. This model passed the consistency ratio test, demonstrating its objectivity and scientific validity. Using this evaluation matrix, we derived the best refugee allocation scheme.

For the third and fourth questions, we compared the existing refugee allocation scheme with our optimized scheme, analyzing the disadvantages and suggesting improvements. This comparison allowed us to provide informed recommendations for better handling the refugee crisis in Europe. We accomplished the above models in the MATLAB, obtaining scientific and reliable data. We compared the similar historical conditions with the reality allocation scheme, obtaining a better result. In addition, we analyzed the sensitivity of the models with different given conditions of data. The fact proved that our models were stable

## 1.1 Symbol and Noun Explanation

In this paper, we use symbols to build our model as shown below:

| symbol | explain |
|---|---|
| x | Europe's parameters and x in different sections have different expression forms and meanings which will explain carefully behind. |
| FN | Refugees accepted index of the Factor Analysis Method |
| F | Refugees accepted index of the Analytic Hierarchy Process |
| β | The factors weight of the Factor Analysis Method |
| α | The factors weight of the Analytic Hierarchy Process |
| CI | coincidence indicator |
| RI | randomly identical target |
| CR | consistency ratio |

## 1.2 Fundamental Assumptions

· Because the refugees' fleeing is lagging. The number appeared to largely fluctuate after 2015. So, we only used the statistical data of 2015 year for analysis.
· Only considering the really number of refugees entering in Europe, we ignored the refugees' loss caused by the factors such as road death.
· In the study of this period of time, we assumed that refugees would remain in Europe, not leaving and not dying
· We assumed that refugees once entered Europe, they could move unimpededly, not having barriers from European countries.
· Europe's development is relatively stable, and its refugee allocation process is static, namely distributing the refugees by percentage.
· The data we collected is reliable and credibly scientific.

## 2. Models

## 2.1 Data Collection

Data collection is consisting of two parts, one is about the data of refugees fleeting to Europe's. The other is the basic data from every European Union country. In terms of the former, we mainly rely on the United Nations refugee agency's official website. As for the latter, we mainly rely on Eurostat's official website. The World Bank, BBC and Reuters also gave us a certain quantity of data.

When we were searching for the data of the refugees, we found that compared to the previous years, the data in 2015 had large fluctuation. As the assumption one described, this fluctuation might be caused by the hysteresis of the Syria war and refugees' psychological factors (Refugees are attached to their native land and unwilling to leave). In 2015 the situation in the Middle East continued unrest, so we can think the number of refugees started from 2015, the refugee crisis also occurred from 2015.

As for the basic data of the European countries, we widely browsed many authoritative data websites. After careful selection, we got the data we need for our models. Of course, we also dealt with the original data.

## 2.2 Prediction Models

Here, we gave two kinds of models. By comparison we found that the second Verhulst model is more aligned with the actual situations. In the end of this section, we also compared similar historical events with the result we predicted, verifying this model.

### 2.2.1 Model One: Logistic Regression Model

The Logistic Regression Model is a kind of widely used models in population growth and population forecast. The refugees themselves have the characteristics of the population: occupying a certain space; existing in a certain period of time; having the same properties as individual collection. Therefore, using Logistic regression model to count and to predict the future of the total number of refugees in Europe is reasonable.

The general form of the Logistic regression model is as follows:

$$f(t) = \frac{a}{a + be^{-kt}} \quad t > 0, \ f > 0$$

Including a, b, k as unknown statistical parameters, t is time, here we used month as unit, f is the total number of refugees entering in Europe.

In order to avoid the influence of the data a few years ago, also in order to simplify the model, we can only use the data from 1 to 10 of 2015 year. The data originated from the UN refugee agency (UNHCR).

In the algorithm, we used the fitting toolbox in MATLAB, using the way of self-defining functions to fit the Logistic regression model, getting the following image:

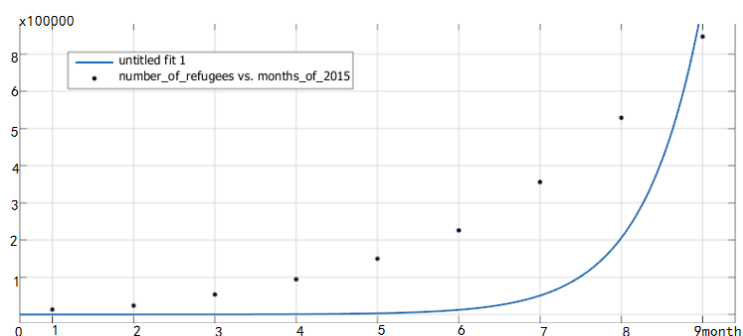

Figure 1. The curve of the number of the refugees based on the Logistic regression fitting

As we can see, the fitting effect is not very good, the value of $R^2$ is 0.492 through the calculation, and every parameter (a, b, k) has too large range of variation, we are unable to get to the limit, so we cannot predict the number of refugees through the regression analysis method.

After analysis, we found that this was due to the lack of statistical data. But we cannot rely on adding statistical data to achieve a better fitting. Because it violates the basic hypothesis one. For this uncertainty problem caused by the poor information, we turned to model two, namely Grayscale Verhulst model.

### 2.2.2 Model Two: Gary Verhulst Model

The Gray Verhulst Model, also known as the Gray Logistic Model, is a common prediction model widely used in population prediction and disease spread. The term "gray" refers to a concept where the reference information is between white (completely transparent) and black (completely unknown), indicating a degree of uncertainty. This model is developed based on the theory of ordinary differential equations. The main implementation process of the Gray Verhulst Model is as follows: First, using the basic idea of difference equations, we perform a first-order cumulative reduction on the original data. Here, $x^{(1)}$ represents the original data, $n$ represents the number of original data points, and $x^{(0)}$ represents the data after the cumulative reduction operation. The result is as follows:

$$x^{(0)}(k) = x^{(1)}(k) - x^{(1)}(k-1), k = 2,3,4 \cdots n$$

Next, we generate the neighbor mean from the original data series $x^{(1)}$, creating the neighbor mean vector $z^{(1)}(k)$:

$$z^{(1)}(k) = \frac{x^{(1)}(k) + x^{(1)}(k-1)}{2}, \quad k = 2,3,4 \cdots n$$

Next, define the data matrices B and Y, and use them to calculate the basic parameter estimate $\tilde{u} = (\tilde{a}, \tilde{b})^T$, the calculation expressions for the data matrices and the estimate are as follows:

$$B = \begin{bmatrix} -z^{(1)}(2) & z^{(1)}(2)^2 \\ \vdots & \vdots \\ -z^{(1)}(n) & z^{(1)}(n)^2 \end{bmatrix} \quad Y = \begin{bmatrix} x^{(0)}(2) \\ \vdots \\ x^{(0)}(n) \end{bmatrix}$$

$$\tilde{u} = (\tilde{a}, \tilde{b})^T = (B^T B)^{-1} B^T Y$$

Finally, we can use the knowledge of ordinary differential equations to solve the differential equation of the Verhulst model. By substituting the above estimated values, $\tilde{a}, \tilde{b}$ into the solution, we obtain the following whitening equation:

$$\tilde{x}^{(1)}(k+1) = \frac{\tilde{a} x^{(0)}(1)}{\tilde{b} x^{(0)}(1) + (\tilde{a} - \tilde{b} x^{(0)}(1)) e^{\tilde{a}k}}$$

We implemented this algorithm through programming in MATLAB. The function and the resulting graph are shown below:

$$g(t) = -\frac{6775.147}{-0.409e^{(-0.502t)} - 0.003}$$

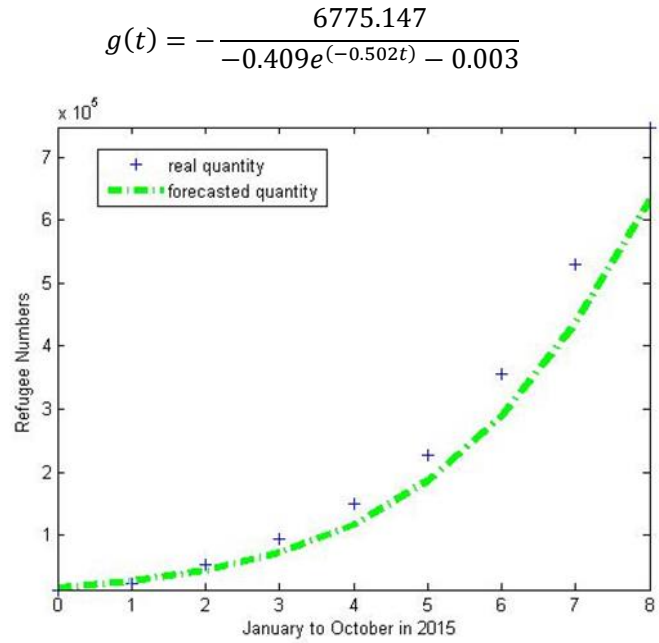

Figure 2. Refugee Population Curve Based on the Gray Verhulst Model

Evidently, the prediction function has an extremum. Without considering other factors, we can roughly estimate the total number of refugees ultimately arriving in Europe. When t is sufficiently large (in fact, the function saturates when t=34t), we obtain $g(t)$ =2003570, approximately two million, which is a significant number.

### 2.2.3 Preliminary Test of the Models

For Model One, we did not adopt it, so we focus on the preliminary verification of Model Two here. (This verification is for accuracy, not stability.) The verification is divided into two aspects: statistical verification of the prediction function itself and verification of the final prediction results.

The Gray Verhulst Model has three commonly used statistical verification methods: relative residual test, variance ratio C test, and small error probability test. According to our calculations, the relative residual Q of the prediction function compared to the original data is 0.017, the variance ratio C is 0.57, and the small error probability p is 0.73. These results indicate that the accuracy level of the model is between Level II and Level III. Therefore, this model has a high degree of reliability and can be used to predict the final number of refugees.

## 2.3 Refugee Distribution Models

Here, we also used two different models to allocate the number of refugees. Through comparative evaluation, we ultimately adopted the Analytic Hierarchy Process (AHP) for refugee allocation.

## 2.3.1 Model One：Refugee Allocation Model Based on the Factor Analysis Method

Factor analysis is an important and practical method in statistics, which is a branch of multivariate statistical analysis. It simplifies a set of variables that reflect the nature, state, and characteristics of a phenomenon into a few inherent factors that can reveal the intrinsic connections of the phenomenon. Simply put, it constructs a multivariate linear equation based on the various factors of the event to predict its future trend.

For the event of refugee population distribution, our research identified four main factors influencing the allocation strategy:
1. Domestic GDP per capita.
2. Domestic average population density.
3. Domestic unemployment rate.
4. Historical number of refugees accepted.

For the last indicator, we use the World Bank's definition of the Public Welfare Index to evaluate the historical number of refugees accepted. This will be specifically explained in Model Two later. Based on these four factors, we constructed a linear regression model for refugee distribution as follows:

$$FN^i = \beta_0 + \beta_1 x_1^i + \beta_2 x_2^i + \beta_3 x_3^i + \beta_4 x_4^i$$

$FN^i$ represents the proportion of refugees that the $i^{th}$ country in the EU should accept. $\beta$ are the weights of each factor for accepting refugees, which are numbers less than 1. x are the corresponding processed data for each country. By fitting and comparing historical data, we obtained these four weights. Substituting these weights into the data, we can obtain the following table:

| Country | FN Value | Rank |
|---|---|---|
| Denmark | 2.346730 | 1 |
| Spain | 2.170950 | 2 |
| Germany | 2.096504 | 3 |

Table 1. Refugee Allocation Chart Based on Factor Analysis Method (Top Three Only)

It can be seen that this allocation scheme differs significantly from the current EU allocation scheme and cannot reasonably solve the refugee distribution issue. While the factor analysis method is simple and easy to understand, determining and predicting various factors is inherently challenging, and the determination of weights often carries subjectivity and is difficult to verify. Additionally, there may be correlations among the various economic and social variables, which can affect the accuracy of the allocation scheme and significantly reduce the credibility of the factor analysis method. Therefore, we adopted Model Two, using the Analytic Hierarchy Process (AHP) to achieve refugee distribution.

## 2.3.2 Model Two: Refugee Allocation Model Based on Analytic Hierarchy Process

As one of the representative discrete models, the Analytic Hierarchy Process (AHP) has always been a powerful tool for handling decision-making problems. Since the issue of refugee resettlement and distribution is discrete in nature, we also adopted AHP here to determine the optimal refugee allocation scheme.

We set the total number of refugees to be allocated as the target layer, with four indicators as the criteria layer, and the scheme layer determined by the specific data of each country. The hierarchy diagram is shown below, and we will explain the reasons and basis for choosing these four criteria in detail next.

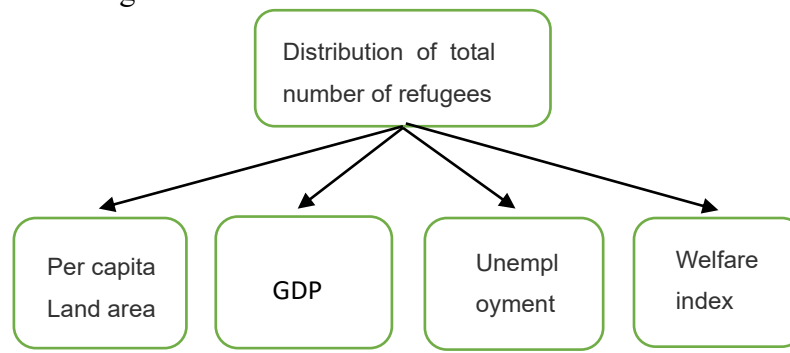

Figure 3. Refugees' allocation scheme based on analytic hierarchy process

1) Land area per capita: Considering man-land relation that, on the one hand, the immigration of refugees will definitely cause the increasing of population density, and on the other hand, the area per capita is an important indicator of population density, the larger the land area per capita a country has, the more refugees the country should accept, and vice versa. Data sources: European statistics agency.

2) GDP: We can make use of GDP (Gross Domestic Product), as an important indicator of state financial resources, to estimate a country's bearing capacity of refugees. Thus, a considerably wealthier country should have the responsibility and capability to accept more refugees. Data sources: European statistics agency.

3) Unemployment rate: When assessing national economy level, unemployment rate is also a significant factor that needs consideration. Accepting refugees means the country has to guarantee their basic life security. However, this could cause demotic resentment, if it is in a high unemployment rate country, because the most important thing for the country is to guarantee the national employment rather than the living standard of refugees. Therefore, the unemployment rate can serve as one of the main criteria. In addition, since the unemployment rate is inversely proportional to the number of admission of refugees, we will apply the rank of unemployment rate to indicate a country's unemployment situation. Data sources: European statistics agency.

4) Public welfare index: Philanthropy Index from Eurostat. The definition of Philanthropy Index stems from Charities Aid Foundation, with the World Bank possessing a similar definition. The Philanthropy Index is used to measure a country's public welfare level, based on specific data, for example, the amount of grain and development funds that have been donated and so on. If a country has already made great contributions in philanthropy, then the country can properly accept less refugees. Hence, we can take Philanthropy Index as one of the main criteria as well. Data sources: the World Bank

The key to the Analytic Hierarchy Process (AHP) lies in the construction of the pairwise comparison matrix. Generally, the comparison matrix is derived from multiple evaluations by experts. For our case, adhering to the principles of objectivity and fairness, each of our three team members independently constructed a comparison matrix. By comparing each other's matrices and referencing literature, we obtained a relatively objective and fair comparison matrix and eigenvalues (weights) as shown below:

$$A = \begin{bmatrix} 1 & \frac{1}{2} & \frac{1}{4} & 2 \\ 2 & 1 & \frac{1}{2} & 3 \\ 4 & 2 & 1 & 5 \\ \frac{1}{2} & \frac{1}{3} & \frac{1}{5} & 1 \end{bmatrix}$$

| Indicator | Land Area per Capita | GDP | Unemployment Rate | Public Welfare Index |
|---|---|---|---|---|
| Weight α | 0.1428 | 0.2641 | 0.5068 | 0.0863 |

Table 2. Weights of Each Criterion in the Criteria Layer

Among them, the comparison factors from left to right (from top to bottom) are Land Area per Capita, GDP, Unemployment Rate, and Public Welfare Index. We also performed a consistency check on this and obtained a CI (Consistency Index) of 0.007, calculated as follows:

$$CR = \frac{CI}{RI} = \frac{0.007}{0.90} = 0.0078 \ll 0.1$$

It can be seen that the comparison matrix passed the consistency check, making it relatively scientific and accurate.

Additionally, because the magnitudes of the original data were not the same, we normalized the original data. Thus, we obtained the processed data as shown below (limited to three countries due to space constraints):

| Country | Relative GDP | Relative Land Area per Capita | Relative Unemployment Rate (%) | Relative Public Welfare Index |
|---|---|---|---|---|
| Ireland | 0.491458621 | 0.061869 | 0.41602317 | 0.928571429 |
| Estonia | 0.548413195 | 0.006428 | 0.26447876 | 0.285714286 |
| Austria | 0.46310015 | 0.117199 | 0.19208494 | 0.821428571 |

Table 3. Relative Indicator Data for EU Countries

Based on this, we can multiply each data point by the weight of each criterion to evaluate and obtain the refugee acceptance index N for each country. The calculation formula is as follows:

$$N^i = \alpha_1 x_1{}^i + \alpha x_2{}^i + \alpha_3 x_3{}^i + \alpha_4 x_4{}^i$$

Here, α represents the weight of each criterion, and x represents the corresponding effective data for each country. Using this formula, we can calculate the number of refugees that each of the 28 EU countries should accept and express it as a proportion. The table below shows the results (only the top and bottom five countries are listed due to space constraints):

| Country | Final Refugee Index | Refugee Acceptance Ratio |
|---|---|---|
| Germany | 0.888911551 | 0.073458167 |
| UK | 0.817150686 | 0.067527969 |
| Austria | 0.656658035 | 0.054265124 |
| Netherlands | 0.630053139 | 0.05206654 |
| Denmark | 0.620816634 | 0.05130325 |
| ⋮ | ⋮ | ⋮ |
| Bulgaria | 0.242810384 | 0.020065445 |
| Slovakia | 0.227811304 | 0.018825946 |
| Cyprus | 0.187564695 | 0.015500033 |
| Croatia | 0.178902624 | 0.014784214 |
| Grace | 0.131674878 | 0.010881392 |

Table 4. Final Proportion of Refugees Each EU Country Should Accept

From the data statistics, we can see that Germany should accept the largest number of refugees, while Greece should accept a smaller number. This aligns with common sense: Germany, as a traditional European power, has a large land area, a high GDP, and a consistently low unemployment rate, giving it the capacity and responsibility to accept more refugees.

In contrast, Greece is facing a severe economic crisis, with a stagnant domestic market and a high unemployment rate, making it difficult to ensure a high standard of living for its people. Therefore, Greece can focus on its economic development and appropriately reduce the number of refugees it accepts.

# 3. Model Examination

## 3.1 History Examination of the Number of Refugees

For the final prediction of 2,003,570 refugees, we validated it through similar historical cases. When the Iraq War broke out in 2003, countless Iraqi refugees were displaced. According to statistics, Iraq's total population in 2003 was approximately 25.96 million (data from the World Bank), and the total number of refugees to date is about 2.5 million (data from IRIN's official website), a ratio of 0.093.

In our case, Syria's population in 2014 was 23.3 million (World Bank). Considering the continued population growth in Syria and the proportion of Syrians among the total number of refugees, the figure of 23.3 million can be compared to the predicted total number of refugees. The ratio is 0.086, which is close to 0.093. Therefore, this approximately confirms the accuracy of our predicted data.

## 3.2 Refugee Allocation Schemes

According to the allocation scheme determined by our Analytic Hierarchy Process (AHP) model, we found that the unemployment rate, which accounts for nearly 50% of the weight, became the primary indicator for evaluating the number of refugees each European country should accept. This is the main reason why Spain ranks so low. Similarly, the public welfare index only accounts for 8% of the weight, so even though Ireland has a high public welfare index (ranking third in that category), its overall ranking is still lower.

This AHP-based allocation scheme can effectively utilize data from various sources and distribute the large number of refugees more evenly among the member countries.

### 3.2.1 Existing Refugee Allocation Scheme

According to Reuters, the final refugee allocation scheme reached by the EU is based on the economic development status and the historical number of refugees accepted by each country. We did not find the specific allocation principles, but we did find the specific allocation numbers as shown in the figure:

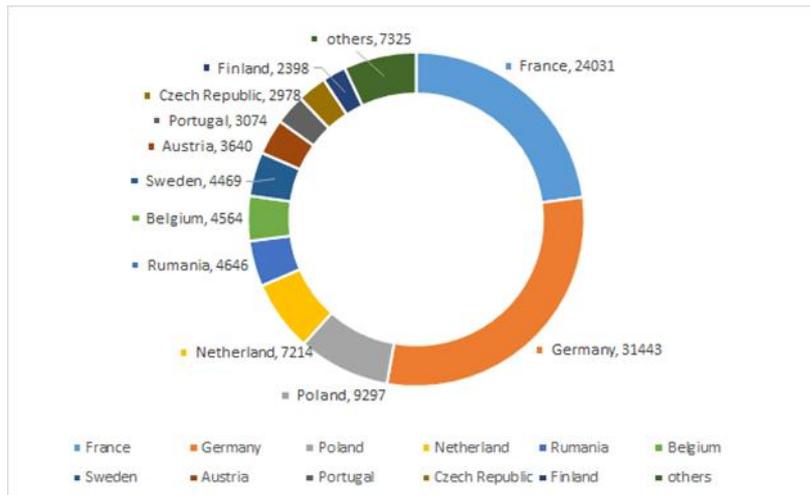

Figure 4. Current Refugee Allocation Scheme in the EU

It can be seen that Germany, France, and Poland are the top three countries in terms of the number of refugees accepted. Germany and France, as traditional European powers, indeed should accept a larger number of refugees. Poland, as an emerging European country, has maintained a good development trend in recent years and also has the capacity to accept refugees.

## 3.2.2 Disadvantages of the Existing Allocation Scheme

Since we could not find the principles of the existing allocation scheme, we cannot analyze its scientific validity based on the principles. However, we can find clues from the specific numbers of refugee allocations to evaluate the current scheme. According to our summary, the existing scheme has the following four shortcomings:

1. It does not fully leverage the collaborative capacity of member countries, nor does it actively mobilize each country's ability to accept refugees. From the figure, we can see that the proportion attributed to "others" is very small, whereas, according to our allocation model, every European country should share the burden of the refugee crisis.
2. The scientific validity of the allocation scheme is insufficient. According to our model, the distribution proportions of countries differ significantly from the actual allocation scheme. The real allocation model may involve political bargaining, leading to uneven distribution. This explains why countries like France and Spain accept so many refugees.
3. The implementation by countries is inadequate, making the allocation scheme difficult to achieve. EU countries are not united, and Eastern European countries consistently oppose the scheme. According to the latest BBC news, only 272 refugees have been successfully resettled under the plan, far from the EU's original goal of 160,000.

4. The UK's overall strength is not much lower than that of Germany, but it has not fulfilled its responsibilities and obligations. The UK has used the "Brexit" as an excuse to bargain with other EU countries. Even though Cameron proposed to accept 20,000 refugees, this is still a drop in the ocean compared to the total of 180,000 refugees. The UK should accept more refugees.

## 3.3 Sensitivity Analysis

Since we ultimately used only two models, we performed a sensitivity analysis on these two models.

## 3.3.1 Refugee Prediction Model

The Gray Verhulst Model itself has good stability. First, we removed one data point (225,146 in May), then we changed the value of another data point (also in May), and finally, we reflected these changes simultaneously in the graph:

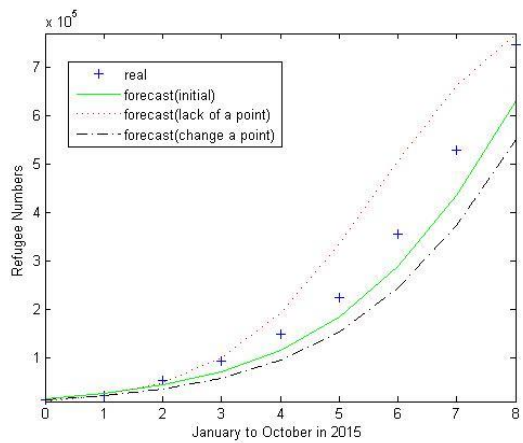

Figure 5. Sensitivity Analysis of the Prediction Curve

From the graph, we can clearly see that our model is relatively stable and not affected by individual data points.

At the same time, we obtained two Gray Verhulst model functions, both of which have extremum points. Therefore, we can use them to predict the final number of refugees, which are 189,398 and 199,590, respectively. Compared to the initial prediction of 2,003,570, these results are still acceptable, indicating the stability of our model.

## 3.3.2 Refugee Allocation Model

For this model, the final results are influenced by both the basic data of each country and the comparison matrix. Therefore, we analyzed the stability of our model from two aspects.

First, we modified the comparison matrix by changing the ratio of the unemployment rate to the public welfare index from 5 to 3, which can be understood as a difference in subjective interpretation. After calculation, we found that the matrix still passed the consistency check, and the weights of each indicator are shown below:

| Indicator | Land Area per Capita | GDP | Unemployment Rate | Public Welfare Index |
|---|---|---|---|---|
| Weight α | 0.1498 | 0.2741 | 0.4717 | 0.1045 |

Table 6. Weight Vector After Changing the Ratio Between Unemployment Rate and Public Welfare Index

From this table, we can see that the weight of the public welfare index has increased, but the overall ranking of the indicators has not changed. The primary factor affecting the refugee allocation scheme remains the unemployment rate. Therefore, we can conclude that our allocation model is relatively stable despite the presence of subjective factors in the comparison matrix.

Next, we choose France as a case study. According to our scheme, France, with its high unemployment rate, should accept fewer refugees. However, in reality, France has the second-highest number of accepted refugees in Europe. Here, we reduce France's unemployment rate by half and observe the impact of this change on the overall allocation scheme:

| Country | Refugee Index |
|---|---|
| Germany | 0.888911551 |
| UK | 0.817150686 |
| France | 0.691911379 |

Table 7. Refugee Acceptance Plan After Reducing France's Unemployment Rate

From this table, we can see that France's ranking has significantly improved, rising from tenth to third place. However, this does not mean that the model is unstable or easily influenced by data points. Our model is designed to change with variations in data points. In this scenario, because France has "resolved" its unemployment issue, it has the capacity to accept more refugees without causing public discontent.

In summary, this model is relatively stable and can be used to address the European refugee crisis effectively.

## 3.4 Applicable conditions of the Models

Based on the final refugee population prediction curve we obtained, we predict that by October 2017, when t=34, the number of refugees in Europe will reach saturation. Therefore, our prediction model is only applicable within this time frame; once t>34, this prediction model loses its specific significance. This is also in line with common sense: the influx of refugees cannot be endless.

As for the allocation model, it is only valid under our basic assumptions, namely that the economic and political situation in Europe will not experience significant fluctuations, that historical data can represent future trends, and that the influx of refugees will not significantly impact Europe's population density, per capita GDP, unemployment rate, and other factors. Additionally, our model is merely a suggestion, an idealized allocation model, and does not pertain to the specific practical schemes of European countries. This means that European countries must put aside their differences and make concerted efforts to effectively allocate refugees, thereby resolving the refugee crisis.

# 4．Conclusion

## 4.1 Advantages and Disadvantages of the Models

In the establishment of the model, we summarize the advantages and disadvantages of our model as shown below:

Advantages：1) It is easy to understand our model and is also convenient to implement and our model has strong ductility. Both gray forecast model and analytic hierarchy process (AHP) have a reliable method which have verified through long-term practice. There are also many cases for reference.

2）Grey forecasting model doesn't need a lot of data points which is very suitable for this case the number of refugees in prediction. Because refugee crisis is explosive and case the valid data much less, which is the strength of gray prediction. Through the analysis of the correlation of the raw data, grey forecasting model can be an effective regression prediction model.

3）Facing with a large number of data samples, we adopt the analytic hierarchy process to solve the problem of distribution of refugees. Analytic hierarchy process (A) can extract the weight of all the data and reasonably arrange refugees. In addition, we compare the opinions of our team and get the evaluated matrix which can effectively avoid the influence of individual subjectivity. This also is one of the strengths of our model.

Disadvantages：1）Our models still lack specific data fitting casing the precision of the model is not high. The dates we found are most before October 2015. In the refugee prediction model, we use the month as a unit which causes the loss of precision. In the refugee allocation scheme, we only adopted four factors as the criterion layer which may affect the model accuracy.

2）Model is too idealistic and without considering the influence of the sudden factors on the model Our political and economic situation of European estimate too optimistic and lack of evidence. Our model simplifies many cases which will also decrease accuracy.

## 4.2 Expansion of the Models

As stated earlier, our model has a strong extensibility. We can use the grey Verhulst model in the population prediction, the spread of disease, population migration model and so on. In these cases, as long as confirmed the basic data of relevance, the model can be passed to predict its future direction.

For the refugees' allocation model based on analytic hierarchy process (AHP), its application is wide. From general fuzzy mathematics to the discrete model, analytic hierarchy process (AHP) has good use of space. So, there is not much to do.

## 4.3 Improvement of the model

In the final refugee population forecast, we can go to search for more effective data and make the time span smaller so that we can improve the model accuracy. These data may be found from the European border agency web site. In addition, we can find out more from the history of case to verify our model, such as the Vietnam War refugees, during the Second World War refugees from all over the world, and so on.

We can use the influx of refugees as a feedback variable to adjust each index of the EU in the model for distribution of refugees. For example: refugees can reduce per capita land area, but it would improve the country's public welfare index, which has the type:

$$x^i(t+1) = x^i(t) + F(t) * k_i(t) * \gamma_i$$

Where k means the proportion of refugees of the country in the current time interval, $\gamma_i$ means the influx of refugees to the influence degree of the indicators. In this way, each time phase of the national situation and distribution proportion is no longer a static numerical refugee but are with time changing variables.

This improvement can be realized simply by BP neural network training. We also need to identify the corresponding layer number, transfer function, threshold of neural and feedback data related to the corresponding negative gradient. In this way, our model accuracy will be improved, and closer to the actual situation.

## 4.4 Suggestions

Here, we put forward the following suggestions according to our model, hoping can help to solve the refugee crisis in Europe.

1. EU countries should be fully prepared. The refugee crisis at least until October 2017 according to our model without the influence of special circumstances. The number of refugees is about 2003570 people ultimately which is a big burden for European countries. Therefore, European countries should strengthen infrastructure construction and formulate correct strategies for refugees. What's more, they also should eliminate prejudice between each other and work together to solve the refugee crisis.
2. Allocating refugees scientifically. Don't obstruct refugees stay in the kingdom of their entry, nor exceed the quota of the refugees to developed countries such as German and French. Our allocation of refugees is based on all EU countries. With the full power of the EU 28 countries, it's easy to allocate refugee. As a result, the more demanded that the European Union countries strengthen cooperation, strictly implement scheme. Here also gives refugees distribution contrast figure.

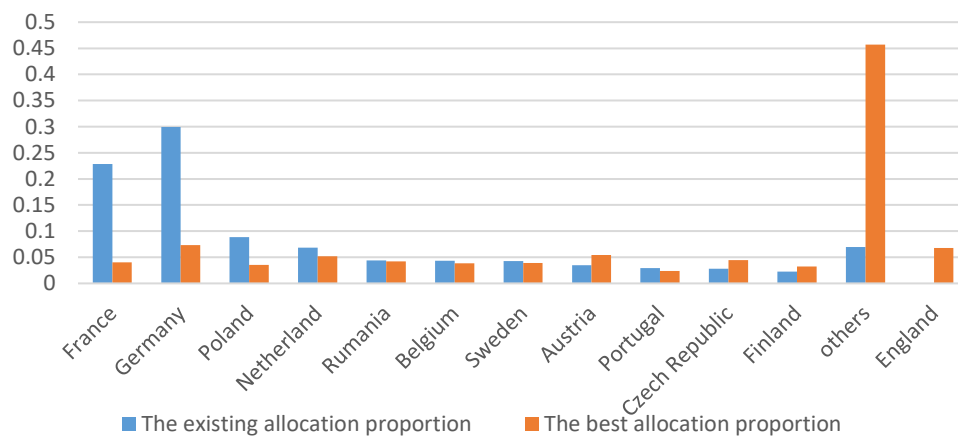

Figure 6. The EU refugee scheme contrast figure

We can see from the figure that the number of the refugees of Germany and France got reduced and Britain also were asked to accept refugees. But the key is that the EU's other countries (others) are to improve the acceptance ratio which is a good way to solve the problem of distribution of refugees.

3. It is turned to face the Britain particular issue. Due to the large population growth rate ranked first in the European Union itself and also accounted for a significant portion of the population structure of immigration. Therefore, the British being reluctant to accept refugees is also for the sake of its own national conditions. But as a European Union member state, accepting refugees was his responsibility and moral duty. This requires the government to adjust outdated ideas and change the refugee policy now.
4. In addition, accepting refugees is not simply to accept them. The EU countries

should follow up some subsequent construction from time to time and make refugees contribute to the country. It also can help the refugees blend in the local humane environment in time which can avoid the emergence of the "country in country".

# 5. References

[1] Jiang Qiyuan; Xie Jinxing; Ye Jun, Mathematical model, Higher Education Press (2011)
[2] Wu Dianting; li Dongfang,The shortage of the AHP and its improving way,Journal of Beijing normal university (natural science edition) ,Apr.2004,Vo1.40 ,No.2
[3] Huang Jianyuan; Liu yang, Prediction Model of Floating Population and Its Applications, STATISTICS AND DECISION, 2008, (23)
[4] S. Liu, J. Zeng, K. Sreenath, and C. A. Belta, "Iterative convex optimization for model predictive control with discrete-time highorder control barrier functions," in 2023 American Control Conference (ACC), 2023, pp. 3368–3375.
[5] UNHCR *http://www.unhcr.org/cgi-bin/texis/vtx/home*
[6] World Bank *http://www.worldbank.org.cn/*
[7] Eurostat *http://ec.europa.eu/eurostat*
[8] Wikipedia *https://en.wikipedia.org/wiki/World_Giving_Index*
[9] Sina blog *http://blog.sina.com.cn/*
[10] Surging news, *http://www.thepaper.cn/*
[11] Google, *http://guge.cytbj.com/webhp?hl=zh-CN*
[12] IRIN, *http://www.irinnews.org/*